# Deep Learning-Driven Quantitative Spectroscopic Photoacoustic Imaging for Segmentation and Oxygen Saturation Estimation

Ruibo Shang [1,2,+], Sidhartha Jandhyala [1,+], Yujia Wu [1,+], Kevin Hoffer-Hawlik [1], Austin Van Namen [1], Matthew O'Donnell [2], Geoffrey P. Luke [1,3*]

[1]*Thayer School of Engineering, Dartmouth College, Hanover, NH 03755, USA*
[2]*uWAMIT Center, Department of Bioengineering, University of Washington, Seattle, WA 98195, USA*
[3]*Translational Engineering in Cancer Research Program, Dartmouth Cancer Center, Lebanon, NH 03766, USA*
+ Equal Contribution

*geoffrey.p.luke@dartmouth.edu

## Abstract

Spectroscopic photoacoustic (sPA) imaging can potentially estimate blood oxygenation saturation ($sO_2$) *in vivo* noninvasively. However, quantitatively accurate results require accurate optical fluence estimates. Robust modeling in heterogeneous tissue, where light with different wavelengths can experience significantly different absorption and scattering, is difficult. In this work, we developed a deep neural network (Hybrid-Net) for sPA imaging to simultaneously estimate $sO_2$ in blood vessels and segment those vessels from surrounding background tissue. $sO_2$ error was minimized only in blood vessels segmented in Hybrid-Net, resulting in more accurate predictions. Hybrid-Net was first trained on simulated sPA data (at 700 nm and 850 nm) representing initial pressure distributions from three-dimensional Monte Carlo simulations of light transport in breast tissue. Then, for experimental verification, the network was retrained on experimental sPA data (at 700 nm and 850 nm) acquired from simple tissue mimicking phantoms with an embedded blood pool. Quantitative measures were used to evaluate Hybrid-Net performance with an averaged segmentation accuracy of ≥ 0.978 in simulations with varying noise levels (0dB-35dB) and 0.998 in the experiment, and an averaged $sO_2$ mean squared error of ≤ 0.048 in simulations with varying noise levels (0dB-35dB) and 0.003 in the experiment. Overall, these results show that Hybrid-Net can provide accurate blood oxygenation without estimating the optical fluence, and this study could lead to improvements in *in-vivo* $sO_2$ estimation.

# 1. Introduction

Photoacoustic (PA) imaging uses a short laser pulse to irradiate biological tissue, where light captured by local absorbers generates an acoustic wave [1, 2]. The propagated acoustic wave is then detected by an ultrasound (US) imaging array transducer and the PA image is reconstructed from collected channel data [3, 4]. In PA image reconstruction, the initial pressure is mapped from channel data using the time-of-flight of the acoustic waves to their source (a process known as the acoustic inverse problem), and the optical absorption coefficient image is reconstructed from the reconstructed initial pressure image and the estimated optical fluence distribution (a process known as the optical inverse problem) [3-5]. Resulting images provide contrast based on optical absorption, but with spatial resolution determined by the acoustic receiver. This makes it particularly well suited to visualize vasculature centimeters deep in tissue.

Spectroscopic PA (sPA) imaging uses a tunable laser to acquire PA images at different light wavelengths [6]. The optical absorption of most endogenous chromophores, such as hemoglobin, and exogenous contrast agents varies significantly across the visible to near-infrared spectrum. Thus, after fluence compensation, the absorber concentration can be estimated from a multi-wavelength dataset by solving a set of equations relating the absorber concentration and absorption spectrum to fluence-compensated sPA data [7, 8]. This approach is called linear unmixing (LU) [9-11] and has been applied to image oxyhemoglobin ($HbO_2$) and deoxyhemoglobin (Hb). The relative concentrations of Hb and $HbO_2$ can then be used to estimate blood oxygen saturation ($sO_2$) throughout tissue [11-14]. Furthermore, by combining spectroscopic PA imaging with real-time US imaging in a handheld probe, true molecular sensitivity can be added to clinical US (i.e., real-time PAUS imaging) [15].

For quantitative sPA imaging, the optical fluence distribution at each wavelength must be known and accounted for to calculate absorber concentrations [6]. However, local optical fluence estimates require propagation properties (i.e., optical scattering, absorption, and scattering anisotropy) that are often unknown [6, 16]. In particular, the light intensity at deeper regions depends on the heterogeneous tissue composition in shallower regions. This attenuation also varies as a function of optical wavelength, a phenomenon known as "spectral coloring" [6]. Thus, blood oxygenation measurements can be skewed as a function of tissue composition and imaging depth, making quantitation difficult.

Deep learning (DL) can significantly enhance the state-of-the-art in image reconstruction compared with conventional algorithms [17-25]. In particular, deep convolutional neural networks trained on large image datasets can learn to reconstruct images by optimizing weights in each convolution layer from gradient descent [18]. Recent efforts in quantifying $sO_2$ with sPA imaging have focused on DL to counteract the effects of spectral coloring [26-34]. For example, two separate convolutional neural networks were proposed to produce 3D maps of vascular blood $sO_2$ and vessel positions from multiwavelength sPA images [32]. The use of a long short-term memory network was proposed for PA oximetry to enable wavelength flexibility of network inputs and Jensen-Shannon divergence to select the most suitable training dataset [33].

Here we present a new DL approach to obtain quantitative $SO_2$ from sPA data based on a deep convolutional neural network (Hybrid-Net). The Hybrid-Net jointly estimates vascular $sO_2$ and segments blood vessels from two-wavelength sPA data. We do not focus on designing a better network architecture or predicting more $sO_2$ information (e.g., 3D $sO_2$ map reconstruction), but show how to further improve the accuracy of $sO_2$ predictions using a single ordinary U-Net architecture. We propose a hybrid loss function incorporating both segmentation and $sO_2$. Segmentation helps focus the network on minimizing the loss value of $sO_2$ predictions within blood vessels, improving $sO_2$ prediction accuracy while enabling clear visualization of the vasculature. The Hybrid-Net was first trained on simulated sPA data with varying noise levels generated from three-dimensional Monte Carlo simulations of light transport in tissues with blood vessels. The model was also trained and made predictions on experimental sPA data acquired from a customized PA imaging system as a proof of concept that the idea can be translated to a real system. The results show that the Hybrid-Net can accurately predict $sO_2$ in blood vessels and segment blood vessels from the background in both simulations and experiments.

## 2. Materials and Methods

### 2.1. Simulated data generation

Simulated sPA data were generated from Monte Carlo simulation using the MCXYZ program [35] as shown in Fig. 1 (a). A 38 × 38 × 38 mm volume was constructed with epidermis, dermis, and breast tissue layers of thickness 0.3 mm, 4.7 mm, and 33 mm, respectively. Representative optical properties were selected for each tissue type [36] as shown in Table 1. Between one and three cylinders with random radius ranging from 0.5 to 4 mm were inserted into the volume with a random orientation. The cylinders extend through the entire length of the volume. The cylinders were filled with blood with a randomly selected $sO_2$. A laser beam with a 4 cm diameter circular aperture was applied normal to the skin surface. Tissue parameters were simulated using two different optical wavelengths: 700 nm and 850 nm.

A total of 4000 simulations (2000 for each wavelength) were run, each for one and a half hours, representing $10^8$ photon packets per simulated volume. Finally, a single two-dimensional cross-section from the center of the absorbed energy volume was taken to be the reconstructed sPA image, assuming that the initial pressure distribution was perfectly reconstructed to focus on the optical problem (Some computational approaches have demonstrated good accuracy in reconstructing the initial pressure distribution [37, 38]). The top 50 rows (~14.8mm) of sPA images (above the black dashed lines in Fig. 1 (a)) were masked to remove the strong signal coming from the epidermis (simulated blood vessels are all located below the 50th row). White Gaussian noise was added to the sPA images and the signal to noise ratio (SNR) calculated as the average power of the PA signal inside the blood vessels divided by the variance of the added noise. 80%, 10%, and 10% of all data was randomly sorted into training, validation, and testing datasets, respectively.

Table 1 Optical properties of simulated tissue

|  | Absorption $\mu_a$ (cm$^{-1}$) (700nm / 850nm) | Scattering $\mu_s$ (cm$^{-1}$) (700nm / 850nm) | Anisotropy of Scattering (g) (700nm / 850nm) |
| --- | --- | --- | --- |
| Epidermis | 0.5542 / 0.2933 | 42.59 / 35.17 | 0.9 / 0.9 |
| Dermis | 0.0168 / 0.0369 | 259.45 / 212.31 | 0.9 / 0.9 |
| Breast | 0.0433 / 0.0575 | 119.76 / 99.02 | 0.9 / 0.9 |

2.2. Experimental data acquisition and preprocessing

Experiments were performed using a custom USPA imaging system as shown in Fig. 1 (b). It consists of an ND: YAG second-harmonic pumped optical parametric oscillator (OPO) laser system (Phocus Mobile HE, Opotek Inc.) with nanosecond-level pulse output at a 10-Hz repetition rate, and a Verasonics Vantage 256 US imaging system with a linear array US transducer (L22-14V, Verasonics). This array has 128 elements, a 0.1 mm element pitch, and a center frequency of 15.625 MHz. The pulsed laser is delivered to the sample through the optical fiber as shown in Fig. 1 (b).

A custom-molded optically-transparent ballistic gel phantom (Humimic Medical) with a blood flow channel (to simulate a blood vessel) was used as the imaging sample. Bovine blood (Lampire Biological Laboratories) was diluted by a 1:1:0.1 ratio of blood to phosphate buffer solution (Corning) to intralipid (Sigma-Aldrich). The blood oxygen level was modulated by purging $CO_2$ gas into the blood and reoxygenated by purging $O_2$ gas into the blood as shown in Fig. 1 (b). The relative position of the blood flow channel with respect to the transducer was changed to simulate different blood vessel locations.

For each $sO_2$ value and vessel location, the phantom was imaged by both PA (at 700 nm and 850 nm) and US modes and reconstructed with conventional PA and US reconstruction methods. Acquired PA images served as sPA data input to the network. Ground-truth $sO_2$ measurements were performed using an optical reflectance probe oximeter, and the ground-truth segmentation was obtained by manually segmenting the acquired US images. Then, the measured ground-truth $sO_2$ value was assigned to the pixels within the ground-truth segmentation image to create the ground-truth $sO_2$ image. A total of 410 sPA signals were acquired with varying $sO_2$ values and vessel locations (each consists of two PA images at 700 nm and 850 nm). 328 (80%) of the entire sPA signals were used as the original training dataset, 41 (10%) were used as the validation dataset, and 41 (10%) were used as the testing dataset. Data augmentation was applied to the original 328 training sPA signals with random rotation, spatial shift and flip, resulting in 1,640 augmented training sPA data.

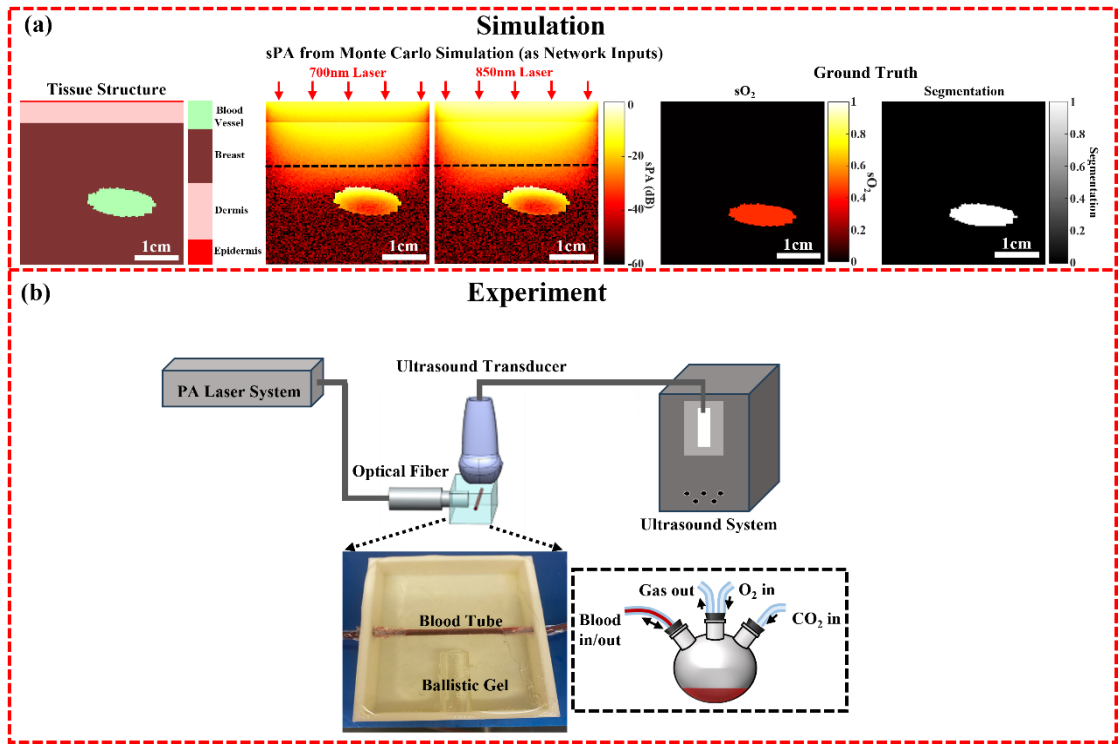

Fig. 1 Data acquisition in simulation and experiments. (a) Data acquisition from Monte Carlo simulation. (b) Experimental setup for sPA data acquisition.

2.3. Hybrid-Net architecture

The Hybrid-Net was based on the widely used U-Net architecture [39], which has an encoder-decoder structure with skip connections between contracting paths and expanding paths, and has been shown to be effective at biomedical image segmentation or other tasks where the output resembles the input [40]. The architecture of the Hybrid-Net is shown in Fig. 2 (c).

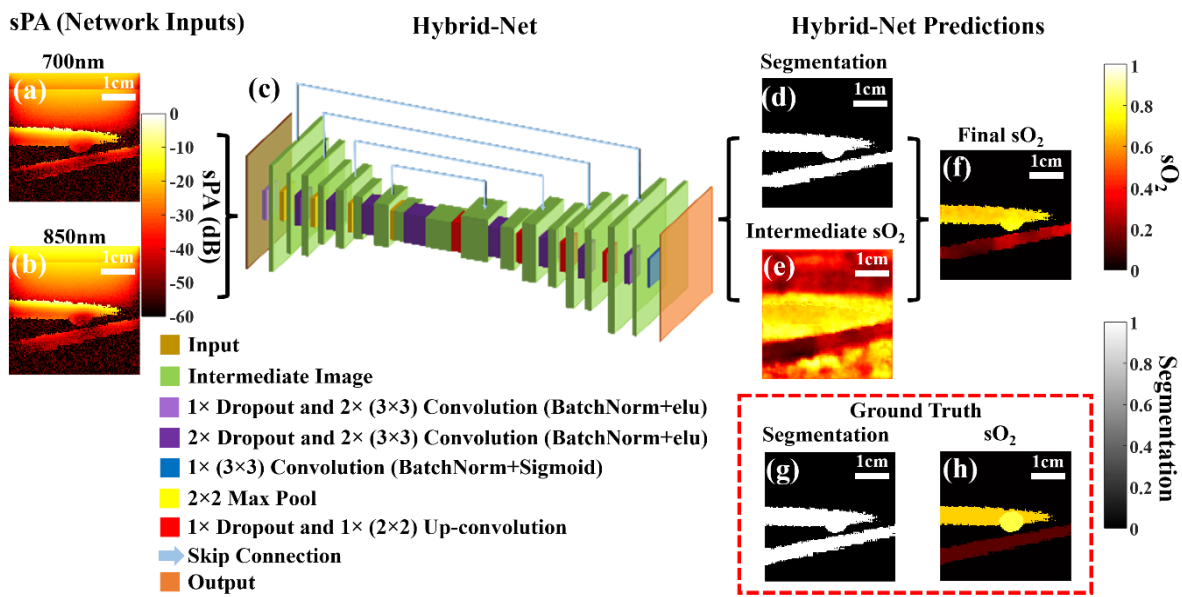

Fig. 2 Hybrid-Net Architecture. (a, b) sPA images at 700nm and 850nm as network inputs. (c) Hybrid-Net architecture based on U-Net. (d) Predicted segmentation image, where a value of 1 represents an artery pixel and a value of 0 represents a background pixel. (e) Predicted sO$_2$ image (intermediate) where a value of 1 represents full oxygenation. (f) Final sO$_2$ image (multiplication of (d) and (e)). (g, h) Ground-truth segmentation and sO$_2$ images.

In general, exponential linear unit (ELU) activation functions were used except for the last layer, where a sigmoid function was applied to predict both the segmentation and sO$_2$ images. Although sigmoid activations are typically used for classification problems, it is appropriate here because both the output segmentation and sO$_2$ are bound between 0 and 1, where for the segmentation image a value of 1 represents a pixel in a blood vessel and a value of 0 represents a pixel in the background, and an sO$_2$ value of 1 represents full oxygenation. Dropout layers with a dropout rate of 0.1 were added before each convolution layer (except the first convolution layer) to avoid overfitting during training [41]. BatchNorm layers were added after each convolution layer (before the activation layer) to stabilize the network and improve network convergence [42]. The network input consisted of the 128×128-pixel, 700- and 850-nm sPA image pair as shown in Fig. 2 (a) and (b).

As stated in *Section 2.1*, we assumed that the initial pressure distribution was perfectly reconstructed to focus on the optical problem. Other computational approaches have demonstrated good accuracy in reconstructing the initial pressure distribution and could be readily combined with the Hybrid-Net [37, 38]. The output was a pair of 128×128 images as shown in Fig. 2 (d) and (e). The first image (Fig. 2(d)) was the predicted segmentation image of the blood vessels. Note that a threshold of 0.5 is applied to the predicted segmentation image directly from the Hybrid-Net. That is, in the final predicted segmentation image, any pixel value larger than 0.5 is set to 1 and all others are set to 0. The second image (Fig. 2 (e)) was the predicted intermediate SO$_2$ map throughout the tissue. Similar to other joint estimation networks [43], a hybrid loss metric was developed for this application as shown in Eq. (1),

$$\text{hybrid loss} = 0.5\text{Dice}(Seg_{pred}, Seg_{gt}) + 0.5\text{MSE}_{seg}(sO_{2_{pred}} \times Seg_{gt}, sO_{2_{gt}} \times Seg_{gt}) \tag{1}$$

where $Seg_{pred}$ is the predicted segmentation image, $Seg_{gt}$ is the ground-truth segmentation image, $sO_{2_{pred}}$ is the predicted sO$_2$ image, $sO_{2_{gt}}$ is the ground-truth sO$_2$ image, Dice denotes the Dice loss function based on the Sørensen-Dice similarity coefficient for measuring overlap between two segmented images [44], and $MSE_{seg}$ denote mean-squared error within the ground-truth segmentation image. Note that MSE of the segmentation was used as the loss function for experimental results.

In Eq. (1), the mean-squared error of $sO_2$ was minimized only in blood vessels (as segmented by the ground-truth segmentation image). While this approach necessarily leads to large errors in background tissue oxygenation estimates, they are neglected by the accompanying vessel segmentation image. The Dice (or MSE) metric was used for segmentation. The two losses ($sO_2$ and segmentation) were weighted equally. Therefore, the final $sO_2$ image as shown in Fig. 2 (f) is the multiplication of the segmentation image (Fig. 2 (d)) and the intermediate $sO_2$ image (Fig. 2(e)).

The hyperparameters in training were chosen based on a previous publication [43]. An Adam optimizer was used with a learning rate of 0.0005. The networks were trained with a batch size of 32 (balance among training speed, overfitting and resistance to fluctuations) and a maximum epoch number of 1,000 to guarantee complete training. An early stopping criterion was applied, where training would finish when the validation loss value did not decrease in 50 consecutive epochs, and optimal weights with the lowest validation loss would be retrieved retrospectively. All training was performed on a NVIDIA GeForce GTX 1080 Ti GPU with 11 GB of memory. Training time was completed in approximately one hour.

To evaluate the perofromance of the hybrid loss function that minimizes $sO_2$ errors only within segmented blood vessels, another neural network, MSE-Net, was trained and compared with Hybrid-Net. In MSE-Net, everything (including training parameters) is the same as Hybrid-Net except that it has only one output channel for $sO_2$ prediction (i.e., no segmentation channel). The loss function for MSE-Net is shown in Eq. (2), which optimize the $sO_2$ in the entire field of view.

$$\text{loss} = \text{MSE}(sO_{2_{pred}}, sO_{2_{gt}}) \tag{2}$$

2.4. Conventional linear unmixing method

Predicted results from Hybrid-Net were also compared with those from conventional linear unmixing, where the $sO_2$ map can be obtained from sPA images by the linear unmixing method. For a fair comparison with Hybrid-Net where the optical fluence is unknown and not estimated, sPA images were used directly as measurements in the linear unmixing method. All tissue was assumed to contain only two optical absorbers – $HbO_2$ and Hb – whose absorption spectrum is known [36]. Then, the PA signal in each pixel was assumed to be proportional to the weighted sum of the two components:

$$PA(\lambda) = \mu_{HbO_2}(\lambda)\, C_{HbO_2} + \mu_{Hb}(\lambda)\, C_{Hb} \tag{3}$$

where $\lambda$ is the optical wavelength, $\mu$ is the optical absorption coefficient of $HbO_2$ or $Hb$, and $C$ is the concentration of $HbO_2$ or $Hb$. After two PA images are acquired using independent wavelengths (700 nm and 850 nm in this paper), Eq. (3) becomes a system of two equations and two unknowns ($C_{HbO_2}$ and $C_{Hb}$) as shown in Eq. (4):

$$\begin{bmatrix} PA_{700}(i) \\ PA_{850}(i) \end{bmatrix} = \begin{bmatrix} \mu_{HbO_2\,700} & \mu_{Hb\,700} \\ \mu_{HbO_2\,850} & \mu_{Hb\,850} \end{bmatrix} \begin{bmatrix} C_{HbO_2}(i) \\ C_{Hb}(i) \end{bmatrix} \tag{4}$$

where $i$ is the pixel number of the image, and the other parameters are the same as Eq. (3).

We solved Eq. (4) using the nonnegative linear least-squares method. Then, the sO$_2$ in each pixel $i$ was calculated as:

$$SO_2(i) = \frac{C_{HbO_2}(i)}{C_{HbO_2}(i) + C_{Hb}(i)} \tag{5}$$

## 3. Results

### 3.1. Simulation results

Hybrid-Net prediction results from two representative images in the simulated testing dataset are shown in Fig. 3. The sPA images at 700 nm and 850 nm are shown in Fig. 3 (a) and (e) as network inputs. White Gaussian noise was added to the sPA images to yield an SNR of 30 dB. The ground-truth segmentation and sO$_2$ images are shown in Fig. 3 (b) and (f). The predicted sO$_2$ images and corresponding errors from Hybrid-Net are shown in Fig. 3 (c) and (g), and the predicted segmentation images and the corresponding errors from Hybrid-Net are shown in Fig. 3 (d) and (h). The segmentation error is the absolute difference between the ground-truth segmentation and predicted segmentation. Note that the sO$_2$ images from Hybrid-Net in Fig. 3 (c) and (g) are the multiplication of predicted (intermediate) sO$_2$ images and the predicted segmentation image as demonstrated in Fig. 2. For comparison, the reconstructed sO$_2$ images and the corresponding errors from MSE-Net and LU are also shown in Fig. 3 (c) and (g). Note that the LU results were only shown within segmented regions to directly compare with those from Hybrid-Net and MSE-Net, and the sO$_2$ error images were only shown within the ground-truth segmented regions for all three approaches to fairly compare errors.

Qualitatively, it can be seen from the sO$_2$ error maps in Fig. 3 (c) and (g) that Hybrid-Net outperforms MSE-Net and LU in sO$_2$ image reconstruction, and segmentation predictions from Hybrid-Net are nearly perfect, with only a few non-zero pixels in segmentation error maps. Hybrid-Net's superior performance to MSE-Net verifies that the hybrid loss minimizing sO$_2$ only within segmented blood vessels in Hybrid-Net leads to more accurate sO$_2$ predictions. In addition, even if the general trends of spectral coloring could be accounted for (e.g., by estimating optical fluence [7]), added noise (PA imaging typically suffers from low SNR in deep tissue) makes LU estimates highly unpredictable.

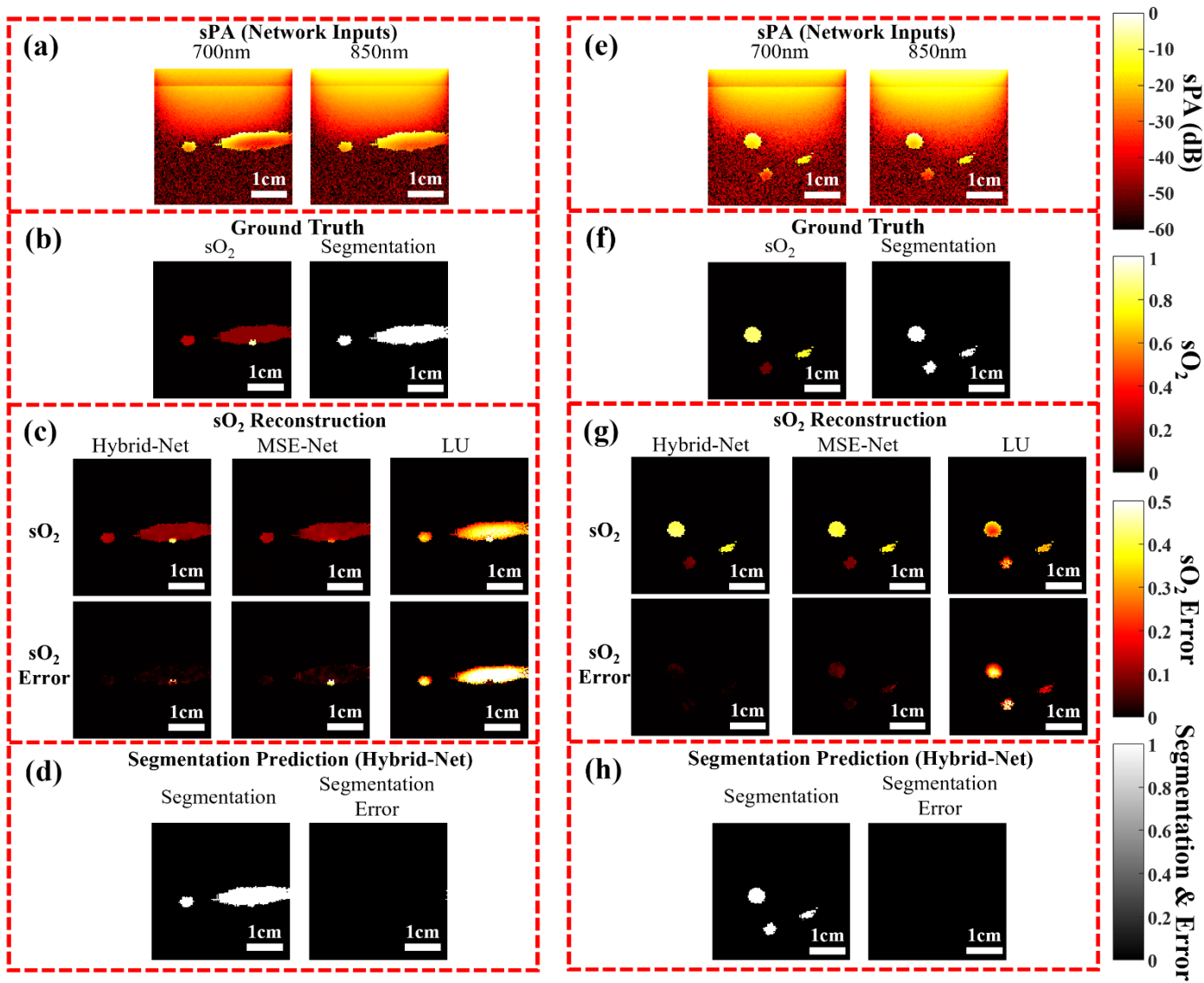

Fig. 3 Simulation results for two samples in the testing dataset. (a) Simulated sPA images with 30-dB SNR at 700nm and 850 nm as network inputs. (b) Ground-truth $sO_2$ and segmentation images. (c) $sO_2$ reconstruction results and corresponding errors from Hybrid-Net, MSE-Net and LU. (d) The predicted segmentation result and corresponding error from Hybrid-Net. (e-h) are the same as (a-d) except for a different sample.

To test Hybrid-Net's performance on $sO_2$ predictions with noisy network inputs, sPA images with varying SNR values of 35, 30, 25, 20, 15, 10, 5 and 0 dB were analyzed. To test the effect of each noise level, Hybrid-Net and MSE-Net were retrained for each noise level. The results for a representative image in the testing dataset are shown in Fig. 4, where again LU results were only shown within segmented regions to directly compare with those from Hybrid-Net and MSE-Net, and the $sO_2$ error images were only shown within the ground-truth segmented regions for all three approaches to fairly compare errors.

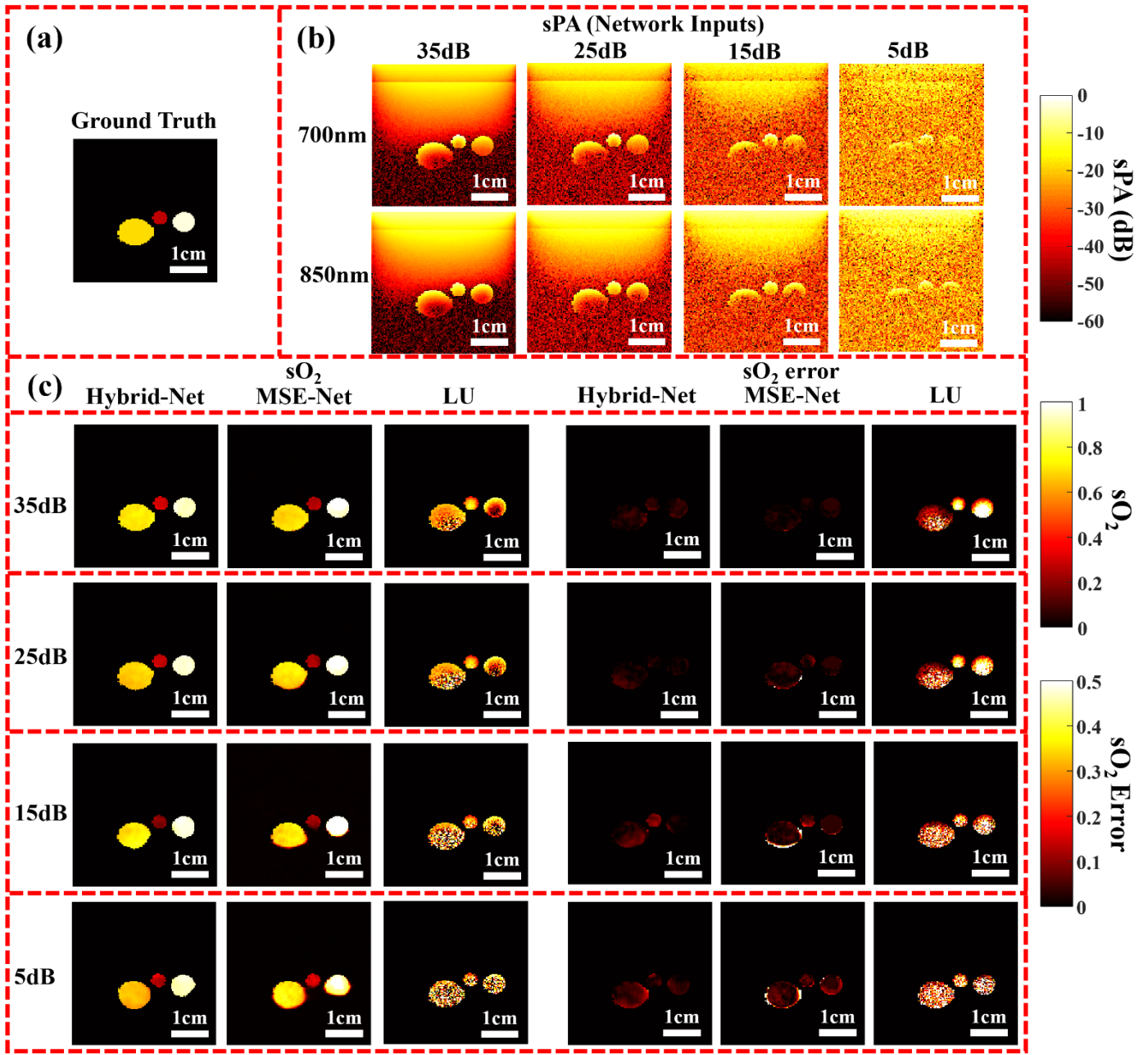

Fig. 4 Simulation results of one sample in the testing dataset with varying noise (SNR) levels. (a) The ground-truth sO$_2$ image. (b) Simulated sPA images at 700 nm and 850 nm as network inputs with varying noise (SNR) levels. (c) sO$_2$ reconstruction results and corresponding errors from Hybrid-Net, MSE-Net and LU.

The ground-truth sO$_2$ image is shown in Fig. 4 (a). sPA images at 700 nm and 850 nm with 35-, 25-, 15- and 5-dB SNR are shown in Fig. 4 (b). As the noise increases (SNR decreases), the PA signal from blood vessels has lower contrast compared to the background. The reconstructed sO$_2$ images and the corresponding errors from Hybrid-Net, MSE-Net and LU are shown in Fig. 4 (c). As the noise level increases (SNR decreases), the performance of both Hybrid-Net and MSE-Net degrades but retains relatively good predictions on the sO$_2$ values. At all four noise levels, Hybrid-Net outperforms MSE-Net with lower sO$_2$ errors, as shown in Fig. 4 (c). Besides, both Hybrid-Net and MSE-Net perform much better than LU. There was little change in LU performance across all noise levels. This is likely the result of a high baseline error arising from spectral coloring. Therefore, qualitatively, Hybrid-Net is more resistant to noise than MSE-Net and LU at varying noise levels.

Quantitative analyses of the performance of Hybrid-Net and MSE-Net are shown in Fig. 5. LU performance is not analyzed here because it does not provide relatively accurate $sO_2$ reconstructions, as shown in Fig. 4. Figure 5 (a) shows the averaged false positive rate, false negative rate and accuracy of segmentation predictions from Hybrid-Net at varying noise levels in the entire testing dataset (MSE-Net is not included because it does not predict segmentation). Overall, the results show Hybrid-Net segmentation predictions are resistant to varying levels of noise with relatively low false positive and false negative rates, and high accuracy. It can also be seen that Hybrid-Net segmentation performance degrades as the noise level increases. This is reasonable because, as the noise level increases, blood vessels are less distinguishable from the background, resulting in increased false negative and false positive rates. Notably, the false negative rate is higher than the false positive rate since the majority of pixels correspond to the background. Hybrid-Net and MSE-Net $sO_2$ performance (within the ground-truth segmented blood vessel regions) in terms of MSE is shown in Fig. 5 (b). The MSE p-values between Hybrid-Net and MSE-Net are less than or equal to 0.0124 for all SNR cases. Clearly, $sO_2$ predictions for both Hybrid-Net and MSE-Net are resistant to varying levels of noise with relatively small MSEs, and Hybrid-Net consistently outperforms MSE-Net at every noise level.

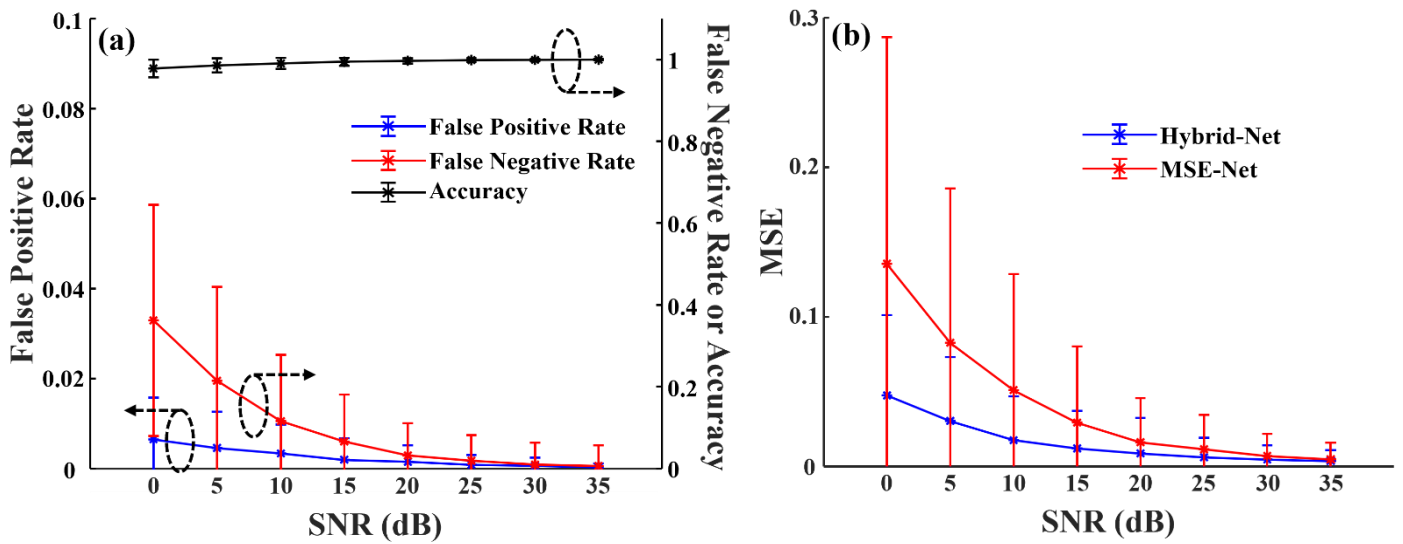

Fig. 5 Quantitative analyses of the performance of Hybrid-Net and MSE-Net for the simulated testing dataset. (a) Averaged false positive rate, false negative rate and accuracy of the predicted segmentation from Hybrid-Net in the entire simulated testing dataset. (b) Averaged MSE of the predicted $sO_2$ from Hybrid-Net and MSE-Net in the entire simulated testing dataset. The error bars denote the standard deviation of the results.

## 3.2. Experimental results

The results of Hybrid-Net, MSE-Net and LU for two samples in the experimental testing dataset are shown in Fig. 6. Figure 6 (a) and (e) show sPA images at 700 nm and 850 nm that are used as network inputs (the sPA signals from the 'blood vessel' are marked by the black dashed ovals). The horizontal lines noted by the white dashed arrows in the sPA images in Fig. 6 (e) are artifacts coming from reverberations of the ultrasound transmission. Ground-truth $sO_2$ and segmentation images for the two samples are shown in Fig. 6 (b) and (f). The predicted (reconstructed) $sO_2$ images and the corresponding errors from Hybrid-Net, MSE-Net and LU are shown in Fig. 6 (c) and (g). Again, LU results were only shown within segmented regions, as in the simulations presented above, to compare with Hybrid-Net and MSE-Net results, and the $sO_2$ error images were only shown within the ground-truth segmented regions for all three approaches to fairly compare errors. Predicted segmentations for Hybrid-Net on the two samples are shown in Fig. 6 (d) and (h).

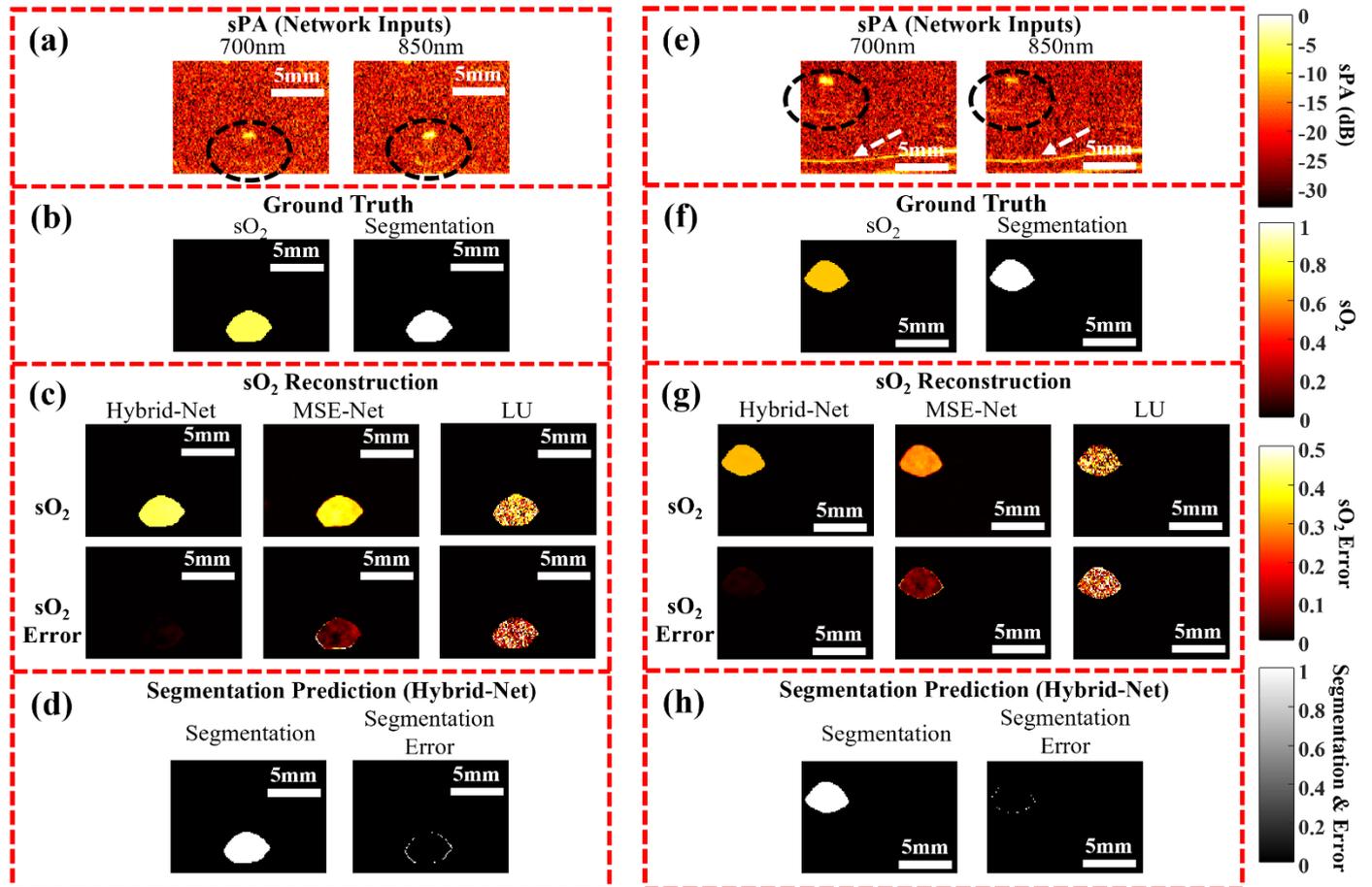

Fig. 6 Experimental results of two samples in the testing dataset. (a) Experimental sPA images at 700 nm and 850 nm as network inputs. (b) Ground-truth $sO_2$ and segmentation images. (c) $sO_2$ reconstruction results and corresponding errors from Hybrid-Net, MSE-Net and LU. (d) Predicted segmentation and corresponding error from Hybrid-Net. (e-h) are the same as (a-d) except for a different sample.

Qualitatively, Hybrid-Net outperforms MSE-Net and LU in sO$_2$ image reconstruction, and segmentation predictions from Hybrid-Net are nearly perfect, with only a few non-zero pixels in segmentation error maps. This experimentally verifies that the hybrid loss function, which minimize sO$_2$ only within segmented blod vessels, leads to more accurate sO$_2$ predictions. Similar to the simulation, deep-learning methods (Hybrid-Net and MSE-Net) also outperform the conventional LU method.

Quantitative evaluations of Hybrid-Net, MSE-Net and LU performance were also conducted. For the segmentation in Hybrid-Net, the averaged false negative rate, false positive rate and accuracy in the experimental testing dataset are 0.0219 ± 0.0114, 0.0009 ± 0.0005 and 0.9982 ± 0.0008 respectively. For the sO2 results, the averaged MSEs in Hybrid-Net, MSE-Net and LU are 0.0027 ± 0.0078, 0.0092 ± 0.0076 and 0.1416 ± 0.0466 respectively. The MSE p-values are 3.4959×10$^{-28}$ between Hybrid-Net and MSE-Net and 1.9898×10$^{-13}$ between Hybrid-Net and LU. Like the findings in the simulation, Hybrid-Net and MSE-Net make relatively accurate sO$_2$ predictions with a small averaged MSE while LU reconstructs inaccurate sO$_2$ with a large averaged MSE. Finally, Hybrid-Net outperforms MSE-Net and LU with a smaller averaged MSE, and Hybrid-Net makes accurate segmentation predictions with small averaged false positive and false negative rates, and high averaged accuracy.

## 4. Discussion

We developed Hybrid-Net to jointly predict vessel segmentation and sO$_2$ for quantitative PA imaging. It was tested using both simulations and proof-of-concept experiments in a simple phantom. Simulations showed that it can accurately predict sO$_2$ and vessel segmentation over a wide range of image SNR. However, as expected, the accuracy of quantitative sO$_2$ estimates decreases as the SNR decreases. There are several straightforward methods to improve the SNR and, hence, improve sO$_2$ accuracy. First, acquiring and averaging multiple PA images at each wavelength increases SNR as the square root of the number of averages at the expense of acquisition time. Similarly, images from more wavelengths can be used as additional network inputs, again at the expense of acquisition time. In addition, the precise wavelengths used for sPA imaging can be optimized to maximize Hb and HbO$_2$ discrimination [8, 45]. Finally, the acoustic inverse problem, vessel segmentation, and sO$_2$ predictions can be combined into a single network, leveraging the results in this paper as well as those in [1, 46-50] that focus on deep learning approaches for the acoustic inversion in PA imaging.

One key benefit of Hybrid-Net is its computational speed. Once trained, blood vessels can be segmented and sO$_2$ estimated in less than 20 ms on a desktop computer. Therefore, the method could be implemented for real-time visualization of vascular oxygenation in clinical applications. Here, we have demonstrated the technique experimentally using a very simple proof-of-concept phantom containing a single blood vessel (tube) and a ballistic gel with similar optical properties at different wavelengths. Clearly, a much more challenging test, including *in vivo* studies, is required to demonstrate clinical applicability. We are currently implementing it on a custom clinical sPA system to image vascular $sO_2$ in the lymph nodes of breast cancer patients suspected of metastatic disease [12].

A major challenge in applying DL to $sO_2$ estimation in vessels is generating a large dataset. Monte Carlo simulation is the most accepted method to accurately estimate light propagation in heterogeneous tissue. The approach, however, is very computationally intensive, relying on stochastic modeling of millions of photon packets, which means building a large dataset is difficult. Therefore, as an additional component of our future work, we will explore how to effectively train Hybrid-Net with smaller datasets that start with acoustic data at individual elements of the imaging array and simultaneously predict sPA images, vessel segmentation and $sO_2$ within segmented vessels.

In this paper, experimental data were used for network training since PA image reconstruction (the acoustic inversion) was not part of the Hybrid-Net model in simulations. This is not practical for clinical applications. If accurate simulations of the complete system can be developed that can be used to train a deep learning model, then models trained on simulation can be directly used on experimental data where ground truth is not available. Such models have been developed for PA image reconstruction (the acoustic inversion) [43, 51] and, based on the results presented in this paper, it seems entirely reasonable that a combined model integrating PA image reconstruction (the acoustic inversion) with vessel segmentation and $sO_2$ estimation can also be developed. In future work, we will explore ways to quantify the prediction error from Hybrid-Net in simulations mimicking practical applications where ground-truth $sO_2$ is unknown (e.g., with a Bayesian convolutional neural network [43, 52-56]).

## 5. Conclusions

We have developed Hybrid-Net to jointly estimate blood $sO_2$ and segment blood vessels from a 700-nm and 850-nm sPA image pair as network inputs. A hybrid loss function was proposed in Eq. (1) to minimize the combined Dice (or MSE) of segmentation and MSE of $sO_2$ within segmented blood vessel regions. Hybrid-Net was tested in both simulations and experiments and compared with the MSE-Net that uses a loss function (in Eq. (2)) to minimize the MSE of $sO_2$ in the entire field of view. Both simulations and experiments show that the Hybrid-Net outperforms MSE-Net, which verifies that the hybrid loss minimizing $sO_2$ only within the segmented blod vessels in Hybrid-Net leads to more accurate $sO_2$ predictions that are robust to noise. The conventional LU method was also applied to the same simulations and experiments, and DL-based methods (Hybrid-Net and MSE-Net) were shown to outperform it. Overall, Hybrid-Net can accurately predict blood vessel segmentation and sO$_2$ in quantitative PA imaging.

## 6. CRediT Authorship Contribution Statement

**Ruibo Shang**: Conceptualization, Data curation, Formal analysis, Investigation, Methodology, Software, Validation, Visualization, Writing – original draft, Writing – review & editing.

**Sidhartha Jandhyala**: Data curation, Formal analysis, Investigation, Methodology, Writing – original draft, Writing – review & editing.

**Yujia Wu**: Data curation, Investigation, Methodology, Writing – original draft, Writing – review & editing.

**Kevin Hoffer-Hawlik**: Investigation, Methodology

**Austin Van Namen**: Investigation, Methodology

**Matthew O'Donnell**: Conceptualization, Methodology, Resources, Writing – review & editing.

**Geoffrey P. Luke**: Conceptualization, Funding acquisition, Methodology, Project administration, Resources, Supervision, Writing – review & editing.

## 7. Declaration of Competing Interests

The authors declare that they have no known competing financial interests or personal relationships that could have appeared to influence the work reported in this paper.

## 8. Data and Code Availability

Data and code underlying the results presented in this paper are not publicly available at this time but may be obtained from the authors upon reasonable request.

## 9. Acknowledgements

This work was supported by the Alma Hass Milham Fellowship in Biomedical Engineering from Thayer School of Engineering at Dartmouth College, a pilot grant from Neukom Institute for Computational Science at Dartmouth College and the National Institutes of Health (R21GM137334 and R01EB030484).